\documentclass[pre,twocolumn,amsmath,amssymb,superscriptaddress]{revtex4}

\usepackage{graphicx}
\usepackage{latexsym}
\usepackage{amsmath}
\usepackage{amssymb}
\usepackage{amsfonts}
\usepackage{bm}

\advance\voffset 1.5cm

\newcommand{\la}{\left<}
\newcommand{\ra}{\right>}

\newcommand{\kB}{\mbox{$k_{\rm B}$}}
\newcommand{\kBT}{\mbox{$k_{\rm B}T$}}

\newcommand{\Ahat}{\ensuremath{\hat{A}}}
\newcommand{\dAhat}{\ensuremath{\delta\hat{A}}}
\newcommand{\Bhat}{\ensuremath{\hat{B}}}
\newcommand{\dBhat}{\ensuremath{\delta\hat{B}}}

\newcommand{\Vhat}{\ensuremath{\hat{V}}}
\newcommand{\Phat}{\ensuremath{\hat{P}}}
\newcommand{\Pid}{\ensuremath{P_\mathrm{id}}}
\newcommand{\Pidhat}{\ensuremath{\hat{P}_\mathrm{id}}}
\newcommand{\Pex}{\ensuremath{P_\mathrm{ex}}}
\newcommand{\Pexhat}{\ensuremath{\hat{P}_\mathrm{ex}}}

\newcommand{\KdvolP}{\ensuremath{\left.K_\mathrm{vol}\right|_P}}

\newcommand{\Krowl}{\ensuremath{K_\mathrm{row}}}
\newcommand{\KrowlV}{\ensuremath{\left.K_\mathrm{row}\right|_V}}
\newcommand{\KrowlP}{\ensuremath{\left.K_\mathrm{row}\right|_P}}

\newcommand{\Hams}{\ensuremath{{\cal H}_s}}
\newcommand{\Uexs}{\ensuremath{{\cal U}_s}}

\newcommand{\etaAex}{\ensuremath{\eta_\mathrm{A,ex}}}
\newcommand{\etaB}{\ensuremath{\eta_\mathrm{B}}}
\newcommand{\etaF}{\ensuremath{\eta}}
\newcommand{\etaFV}{\ensuremath{\left.\eta\right|_V}}
\newcommand{\etaFP}{\ensuremath{\left.\eta\right|_P}}

\newcommand{\etaFidP}{\ensuremath{\left.\eta_\mathrm{id}\right|_P}}

\newcommand{\etaFmixP}{\ensuremath{\left.\eta_\mathrm{mix}\right|_P}}

\newcommand{\NVT}{\ensuremath{\text{NVT}}}
\newcommand{\NPT}{\ensuremath{\text{NPT}}}

\newcommand{\dk}{\ensuremath{\delta k}}
\newcommand{\kl}{\ensuremath{k_l}}
\newcommand{\Rl}{\ensuremath{R_l}}

\newcommand{\xl}{\ensuremath{x_l}}
\newcommand{\fl}{\ensuremath{f_l}}

\begin{document}

\title{Pressure fluctuations in isotropic solids and fluids}

\author{J.P.~Wittmer}
\email{joachim.wittmer@ics-cnrs.unistra.fr}
\affiliation{Institut Charles Sadron, Universit\'e de Strasbourg \& CNRS, 23 rue du Loess, 67034 Strasbourg Cedex, France}
\author{H.~Xu}
\affiliation{LCP-A2MC, Institut Jean Barriol, Universit\'e de Lorraine \& CNRS,\\ 1 bd Arago, 57078 Metz Cedex 03, France}
\author{P.~Poli\'nska}
\affiliation{Institut Charles Sadron, Universit\'e de Strasbourg \& CNRS, 23 rue du Loess, 67034 Strasbourg Cedex, France}
\author{F.~Weysser}
\affiliation{Institut Charles Sadron, Universit\'e de Strasbourg \& CNRS, 23 rue du Loess, 67034 Strasbourg Cedex, France}
\author{J. Baschnagel}
\affiliation{Institut Charles Sadron, Universit\'e de Strasbourg \& CNRS, 23 rue du Loess, 67034 Strasbourg Cedex, France}

\begin{abstract}
Comparing isotropic solids and fluids at either imposed volume or pressure we investigate various
correlations of the instantaneous pressure and its ideal and excess contributions.
Focusing on the compression modulus $K$ it is emphasized that the stress fluctuation representation of the 
elastic moduli may be obtained directly (without a microscopic displacement field)
by comparing the stress fluctuations in conjugated ensembles. 
%
This is made manifest by computing the Rowlinson stress fluctuation expression $\Krowl$ 
of the compression modulus for \NPT-ensembles. It is shown theoretically 
and numerically that $\KrowlP = \Pid (2 - \Pid/K)$ with $\Pid$ being the ideal pressure contribution.
\end{abstract}


\date{\today}
\maketitle

\paragraph*{Introduction.}
Among the fundamental properties of any equilibrium system are its elastic moduli 
characterizing the fluctuations of its extensive and/or conjugated intensive variables 
\cite{BornHuang,Callen,ChaikinBook,RowlinsonBook,Hoover69,Lutsko89,WTBL02}.
The isothermal compression modulus $K$ of an isotropic solid or fluid may thus be obtained 
in the $\NPT$-ensemble at imposed particle number $N$, pressure $P$ and temperature $T$
from the fluctuations $\delta \Vhat = \Vhat - V$ of the instantaneous volume $\Vhat$ 
around its mean value $V = \langle \Vhat \rangle$ according to the {\em strain fluctuation} relation \cite{Callen}
\begin{equation}
K = \KdvolP \equiv 
\kBT V / \left.\langle \delta \Vhat ^2 \rangle\right|_P
\label{eq_K_NPT}
\end{equation}
with $\kB$ being Boltzmann's constant.
Equivalently, $K$ may be obtained in a canonical $\NVT$-ensemble 
using Rowlinson's {\em stress fluctuation} relation \cite{RowlinsonBook,AllenTildesleyBook,WXP13}
\begin{equation}
K = \KrowlV \equiv P + \etaB - \beta V \left.\la \delta \Pexhat^2 \ra\right|_V
\label{eq_Rowl}
\end{equation}
with $\beta = 1/\kBT$ being the inverse temperature, $\Pexhat$ the instantaneous
excess pressure contribution and $\etaB$ a Born-Lam\'e coefficient \cite{WXP13}
which for pairwise additive potentials becomes a simple sum of moments of derivatives of the potential 
with respect to the particle distance \cite{AllenTildesleyBook}.
In this Communication we emphasize that the stress fluctuation representation of the elastic moduli 
\cite{RowlinsonBook,Hoover69,Lutsko89} may be obtained directly from the well-known transformation 
rules between conjugated ensembles \cite{Lebowitz67}. 
Focusing on the compression modulus $K$ this is made manifest by computing Rowlinson's expression $\Krowl$ 
deliberately for \NPT-ensembles where the volume is allowed to freely fluctuate. We show that
\begin{equation}
\KrowlP =  \Pid \ (2 - \Pid/K)
\label{eq_key}
\end{equation}
with $\Pid$ being the ideal pressure contribution.
We demonstrate first Eq.~(\ref{eq_key}) by considering theoretically 
the fluctuations of the instantaneous normal pressure $\Phat$ and 
its ideal and excess contributions $\Pidhat$ and $\Pexhat$ in both conjugated ensembles.
These correlations are then checked numerically by means of Monte Carlo (MC) simulation of simple coarse-grained model systems. 

\paragraph*{Background.}
As discussed in the literature \cite{Callen,AllenTildesleyBook,Lebowitz67} a simple average $A = \langle \hat{A} \rangle$ 
of an observable ${\cal A}$ does not depend on the chosen ensemble, at least not if the system is large enough 
($V \to \infty$). A correlation function $\langle \dAhat \dBhat \rangle$ of two observables ${\cal A}$ and ${\cal B}$ 
may differ, however, depending on whether $V$ or $P$ are imposed. As shown by Lebowitz, Percus and Verlet 
\cite{Lebowitz67} one verifies that
\begin{equation}
\left. \la \dAhat \dBhat \ra\right|_{V} = \left. \la \dAhat \dBhat \ra\right|_{P}
- \frac{K}{\beta V} \ \frac{\partial A}{\partial P} \frac{\partial B}{\partial P}
\label{eq_dAdB}
\end{equation}
where $K = - V \partial P / \partial V$ \cite{Callen} has been used.
For $\Ahat=\Bhat=\Phat$ this implies the transformation 
\begin{equation}
\beta V \left.\la \delta \Phat^2 \ra\right|_V = \beta V \left.\la \delta \Phat^2 \ra\right|_P - K,
\label{eq_Pfluctutrans}
\end{equation}
i.e. the compression modulus $K$ may be obtained from the difference of the 
pressure fluctuations in both ensembles. Interestingly, the numerically more convenient Rowlinson 
expression $\Krowl$ for $\NVT$-ensembles can be derived from Eq.~(\ref{eq_Pfluctutrans}) \cite{WXP13}
without using a microscopic displacement field (only possible for solids) \cite{Hoover69} and 
avoiding the volume rescaling trick used originally for liquids \cite{RowlinsonBook}.

\paragraph*{MC-gauge.}
There is a considerable freedom for defining the instantaneous pressure
$\Phat = \Pidhat + \Pexhat$ as long as its average $P = \Pid + \Pex$ does not change 
\cite{AllenTildesleyBook}.
It is convenient for the subsequent derivations and the presented MC simulations
to define the instantaneous ideal pressure $\Pidhat$ by
\begin{equation}
\Pidhat = \kBT N /\Vhat \ \ \mbox{(MC-gauge)}
\label{eq_MCgauge}
\end{equation}
and the instantaneous excess pressure $\Pexhat$ by the Kirkwood expression \cite{AllenTildesleyBook,WXP13}.
Within this ``MC-gauge" the thermal momentum fluctuations are assumed to be integrated out
and the (effective) Hamiltonian $\Hams$ of a state $s$ of the system may be written
\begin{equation}
\Hams(\Vhat) = - \kBT N \log(\Vhat) + \Uexs(\Vhat) + \text{consts}
\label{eq_Hams}
\end{equation}
with $\Uexs$ being the total excess potential energy.

\paragraph*{Non-affine contribution.}
In the following the concise notation $\etaF \equiv \beta V \langle \delta \Phat^2 \rangle$ is used.
An immediate consequence of the MC-gauge is, of course, that the fluctuations of $\Pidhat$
vanish for the \NVT-ensemble and that, hence, 
\begin{equation}
\etaFV = \beta V \left.\la \delta \Pexhat^2 \ra\right|_V.
\label{eq_etaPV}
\end{equation}
Since $K > 0$ for a stable system, Eq.~(\ref{eq_Pfluctutrans}) implies $\etaFP > \etaFV$.
Depending on the disorder, $\etaFV$ is, however, not a negligible contribution
as assumed (implicitly) 
by Born \cite{BornHuang}. For solids it measures the effect of {\em non-affine} 
displacements under an imposed macroscopic strain \cite{Lutsko89,WTBL02,WXP13}.

\paragraph*{Affine (Born) contribution.}
The second moment of any intensive variable computed in an ensemble, where its mean value is imposed,
is obtained readily by integration by parts. Using Eq.~(\ref{eq_Hams}) this shows that 
\begin{equation}
\etaFP = \left.V \la \Hams^{\prime\prime}(\Vhat)\ra\right|_P = \Pid + \left.V \la \Uexs^{\prime\prime}(\Vhat)\ra\right|_P
\label{eq_etaA}
\end{equation}
where a prime denotes a derivative with respect to the indicated variable.
Albeit the indicated averages are taken over all states $s$
and all volumes $\Vhat$ at imposed $P$, being {\em simple averages}
they can also be evaluated for sufficiently large systems in the $\NVT$-ensemble
yielding identical results.
Denoting the last term in Eq.~(\ref{eq_etaA}) by $\etaAex$ one can show
that it is equivalent for pair interaction potentials to the already mentioned Born-Lam\'e coefficient
\cite{WXP13}: 
$\etaAex = \etaB + \Pex$.
Substituting Eq.~(\ref{eq_etaPV}) and Eq.~(\ref{eq_etaA}) into the Legendre transform
Eq.~(\ref{eq_Pfluctutrans}), this confirms Eq.~(\ref{eq_Rowl}).
\paragraph*{Correlations at constant $P$.}
We focus now on stress fluctuations in the \NPT-ensemble.
By comparing with Eq.~(\ref{eq_Pfluctutrans}) one sees that if the Rowlinson formula $\Krowl$ 
is applied at imposed $P$, this must yield
\begin{equation}
\KrowlP = \beta V\left.\la \delta \Pidhat^2 \ra\right|_P + 2 \beta V\left.\la \delta \Pidhat \delta \Pexhat \ra\right|_P.
\label{eq_KrowlPcorrfunc}
\end{equation}
Interestingly, Eq.~(\ref{eq_KrowlPcorrfunc}) does not completely vanish for finite $T$ as does the corresponding
stress fluctuation expression for the shear modulus $G$ at imposed shear stress $\tau$ \cite{WXP13}.
As a next step we demonstrate the relations 
\begin{eqnarray}
\etaFidP 
\equiv \beta V \left.\la \delta \Pidhat^2 \ra\right|_P  
& = & \Pid^2/K \label{eq_etaFid} \\
\etaFmixP 
\equiv \beta V \left.\la \delta \Pidhat \Pexhat \ra\right|_P 
& = &  \Pid \ (1 - \Pid/K) \label{eq_etaFmix} 
\end{eqnarray}
from which Eq.~(\ref{eq_key}) is then directly obtained by substitution into 
Eq.~(\ref{eq_KrowlPcorrfunc}).
Returning to the general transformation relation Eq.~(\ref{eq_dAdB}) we note first that the
{\em l.h.s.} must vanish if at least one of the observables is
a function of $\Vhat$. With $\Ahat = \Bhat = 1/\Vhat$ it follows that 
\begin{equation}
\left.\la \delta (1/\Vhat)^2 \ra\right|_P 
= \frac{K}{\beta V} \ \left(\frac{\partial \langle 1/\Vhat\rangle}{\partial P}\right)^2
\approx \frac{1}{\beta K V^3}
\label{eq_VinvfluctuA}
\end{equation} 
making the steepest-descent approximation $\langle 1/\Vhat \rangle \approx 1/V$
for simple averages and using finally $V/K = -\partial V / \partial P$ \cite{foot_gauss}.
Remembering Eq.~(\ref{eq_MCgauge}) this implies Eq.~(\ref{eq_etaFid}).
With $\Ahat =  \Pidhat = \kBT N/\Vhat$ and $\Bhat = \Phat$ one obtains similary
\begin{equation}
\beta V \left.\la \delta \Pidhat \delta \Phat \ra\right|_P
= \kBT N K \ \frac{\partial \langle 1/\Vhat\rangle}{\partial P}
 \approx \Pid 
\label{eq_VinvfluctuB}
\end{equation}
to leading order for $V \to \infty$. This relation implies finally the claimed correlation between 
ideal and excess pressure fluctuations, Eq.~(\ref{eq_etaFmix}), using $\Phat = \Pidhat + \Pexhat$ and 
the already demonstrated Eq.~(\ref{eq_etaFid}) \cite{foot_etaFex}.
Please note that in Ref.~\cite{WXP13} ideal and excess pressure fluctuations have incorrectly been
assumed to be uncorrelated.

\begin{figure}[t]
\centerline{\resizebox{0.95\columnwidth}{!}{\includegraphics*{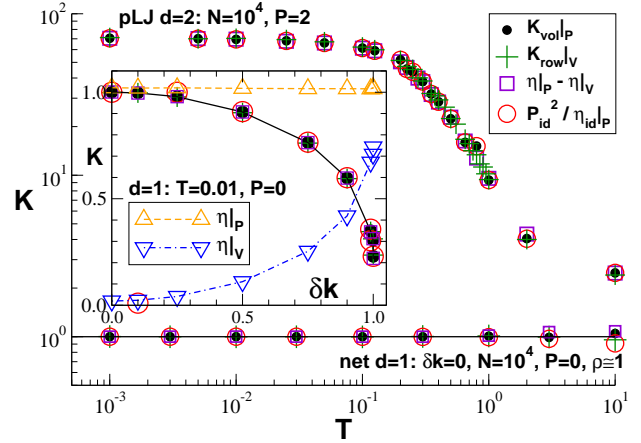}}}
\caption{Compression modulus $K$ computed using 
the rescaled volume fluctuations $\KdvolP$ (filled spheres),
the Rowlinson stress fluctuation formula $\KrowlV$ (crosses),
the difference between the total pressure fluctuations in both ensembles (squares) and
the fluctuations of the inverse volume $\Pid^2/\etaFidP$ (large spheres).
Main panel: 
The upper data refer to systems of glass-forming 2D pLJ beads at $P=2$ \cite{WXP13},
the lower data to simple 1D nets of harmonic springs at $P=0$.
Inset: Compression modulus $K$ {\em vs.} polydispersity $\dk$ of the spring constants
for 1D nets with $T=0.01$ and $P=0$.
\label{fig_K}
}
\end{figure}

\paragraph*{Some algorithmic details.}
The numerical results reported here to check our predictions have been obtained by MC simulation of 
{\em (i)} one-dimensional (1D) nets with permanent cross-links and 
{\em (ii)} two-dimensional (2D) glass-forming liquids.
Periodic boundary conditions are used and the pressure $P$ is first imposed
using a standard MC barostat \cite{AllenTildesleyBook,WXP13}. 
After equilibrating and sampling in the \NPT-ensemble, the volume is fixed, $V=\Vhat$, and various simple means 
and fluctuations are obtained in the \NVT-ensemble at the same state point \cite{foot_volume}.
For the 1D nets we assume ideal harmonic springs, $U = \sum_l \kl (\xl - \Rl)^2/2$,
with $\xl$ being the distance between the connected particles,
the reference length $\Rl$ of the springs being set to unity and the spring constants $\kl$ 
being taken randomly from a uniform distribution of half-width $\dk$ 
centered around a mean value also set to unity. Only simple networks are presented here where two particles
$i-1$ and $i$ along the chain are connected by {\em one} spring $l=i$, i.e. at zero temperatures all forces $\fl$ 
along the chain become identical. This implies $K \sim 1/\langle 1/\kl \rangle$. 
The compression modulus decreases thus strongly with $\dk$
as indicated by the bold line in the inset of Fig.~\ref{fig_K}. 
Our 2D systems are polydisperse Lennard-Jones (pLJ) beads \cite{WTBL02} kept at a constant pressure $P=2$
as described in Ref.~\cite{WXP13}. 

\paragraph*{Computational results.}
As shown in Fig.~\ref{fig_K}, the compression modulus $K$ may be determined using the volume 
fluctuations in the \NPT-ensemble, Eq.~(\ref{eq_K_NPT}), or using Rowlinson's stress fluctuation formula, 
Eq.~(\ref{eq_Rowl}), for the \NVT-ensemble. 
The same values of $K$ are obtained from the Legendre transform for the pressure fluctuations, 
Eq.~(\ref{eq_Pfluctutrans}), and from the ideal pressure fluctuations $\etaFidP$, Eq.~(\ref{eq_etaFid}),
which thus confirms both relations.
As seen in the inset of Fig.~\ref{fig_K}, the compression modulus of the 1D nets decreases with $\dk$.
Also indicated is the ``affine" contribution $\etaFP$ to $K$, measuring the mean spring constant $\langle \kl \rangle =1$,
and the ``non-affine" contribution $\etaFV$ which is seen to increase with $\dk$.
The decrease of $K$ is thus due to the increase of the non-affine contribution.
\begin{figure}[t]
\centerline{\resizebox{0.95\columnwidth}{!}{\includegraphics*{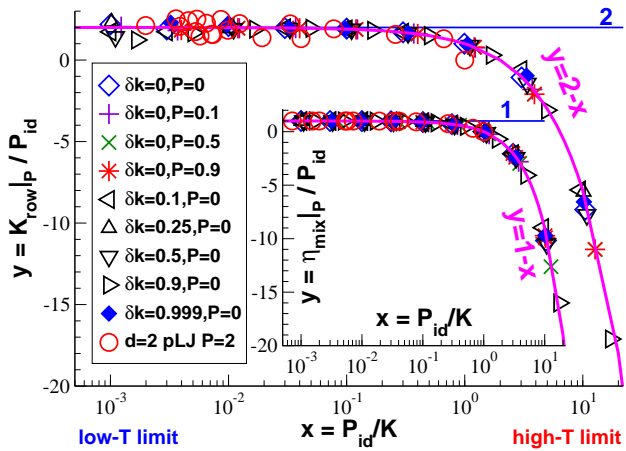}}}
\caption{Characterization of stress fluctuations in the \NPT-ensemble.
Large spheres refer to 2D pLJ beads for $P=2$, all other symbols to 1D nets for 
different $\dk$ and $P$ as indicated.
Main panel:
Rescaled Rowlinson formula $\KrowlP/\Pid$ as a function of the reduced ideal pressure $x = \Pid/K$.
The bold line represents our key prediction, Eq.~(\ref{eq_key}), on which all data points collapse.
Inset: Similar scaling for the reduced correlation function $\etaFmixP/\Pid$ confirming Eq.~(\ref{eq_etaFmix}).
\label{fig_NPT}
}
\end{figure}

As shown in the inset of Fig.~\ref{fig_NPT}, we have also checked the correlations between the
ideal and the excess pressure fluctuations $\etaFmixP$. To make both models comparable the reduced 
correlation function $y = \etaFmixP /\Pid$ is traced as a function of the reduced ideal pressure 
$x = \Pid/K$ with $K$ as determined independently above. A perfect data collapse on the prediction 
$y = 1-x$ (bold line) is observed for all systems.
The main panel of Fig.~\ref{fig_NPT} shows finally the scaling of the Rowlinson formula computed 
in the \NPT-ensemble. 
As before a scaling collapse of the data is achieved by plotting $y = \KrowlP/\Pid$ {\em vs.} $x$.
The bold line indicates our key prediction, Eq.~(\ref{eq_key}).
Interestingly, the latter result does not depend on the MC-gauge which has been used above
to simplify the derivation of Eq.~(\ref{eq_key}).
Please note that it is not possible to increase $x$ beyond unity for our liquid systems ($K \ge \Pid$) 
and the deviations from the low-temperature plateau $y = 2$ are thus necessarily small.
The additional $\Pid/K$ correction has thus been overlooked in our previous publication \cite{WXP13}.

\paragraph*{Conclusion.}
Emphasizing the underlying Legendre transform, Eq.~(\ref{eq_Pfluctutrans}), of the stress fluctuation formalism,
we have investigated here the well-known Rowlinson stress fluctuation expression $\Krowl$ for the compression modulus, 
Eq.~(\ref{eq_Rowl}), using deliberately the \NPT-ensemble.  Correcting several statements made in Ref.~\cite{WXP13}, 
it has been demonstrated theoretically and numerically that Eq.~(\ref{eq_key}) holds.
The latter result, as the other correlation relations indicated in the paper, may allow to readily
calibrate (correctness, convergence and precision) various barostats commonly used \cite{AllenTildesleyBook}.

\paragraph*{Acknowledgements.}
P.P. thanks the  R\'egion Alsace and the IRTG Soft Matter and F.W. the DAAD for funding.
We are indebted to A.~Blumen (Freiburg) and A. Johner (Strasbourg) for helpful discussions.


\begin{thebibliography}{13}
\bibitem[{foo({\natexlab{a}})}]{foot_gauss}
\bibinfo{note}{That Eq.~(\ref{eq_VinvfluctuA}) becomes exact for $V \to \infty$
  can be also seen by using that the distribution of $\Vhat$ is Gaussian.}
\bibitem[{foo({\natexlab{b}})}]{foot_etaFex}
\bibinfo{note}{For the excess pressure fluctuations a similar relation
  $$\beta V \left.\langle \delta \Pexhat^2 \rangle\right|_P = \etaAex - \Pid \
  (1 - \Pid/K)$$ is obtained using Eq.~(\ref{eq_etaA}) together with
  Eqs.~(\ref{eq_etaFid},\ref{eq_etaFmix}). This relation has been also
  confirmed numerically.}
\bibitem[{foo({\natexlab{c}})}]{foot_volume}
\bibinfo{note}{In $d=1$ ``volume" corresponds to the linear length of the
  system and in $d=2$ to its surface. Pressure and elastic moduli take units of
  energy per $d$-dimensional volume.}
\end{thebibliography}

\end{document}